# Nearly ballistic transport and high magnetic-field sensitivity in a $Bi_4Br_4$ topological Josephson weak link

Enamul Haque[1], and Javier Cerrillo[1]

[1]Área de Física Aplicada, Universidad Politécnica de Cartagena, 30202 Cartagena, Spain

Corresponding authors: Enamul Haque—enamul.haque@upct.es
Javier Cerrillo—javier.cerrillo@upct.es

## Abstract

Superconducting weak links with high transparency offer an appealing approach for designing compact magnetic-field sensors, as their phase-dependent Andreev bound-state (ABS) spectrum produces a strong flux-to-signal response with minimal dissipation. One way to achieve ballistic transport in the weak link is to use the edge states of topological insulators, since these states resist backscattering and provide a unique path to developing topological insulator-based weak-link devices for highly efficient magnetic-field sensing. Building on this, we propose a weak-link device using superconducting Nb electrodes and a nanoribbon of $Bi_4Br_4$ as the normal region, forming a Nb—1D (one-dimensional) $Bi_4Br_4$—Nb Josephson junction. We develop first-principles tight-binding Hamiltonians and orbital-resolved interface couplings in the Wannier basis, including spin–orbit coupling, based on density functional theory (DFT) calculations. The Eliashberg spectral function of bulk Nb, obtained via density functional perturbation theory (DFPT), indicates an electron–phonon coupling strength of 1.19 and a transition temperature of about 9 K, aligning well with conventional superconductivity in Nb. Transport in the normal region involves a small number of channels, including two highly transmitting channels and two additional weakly transmitting modes. The subgap conductance is primarily influenced by Andreev processes. The ABS spectrum leads to a non-sinusoidal current-phase relation (CPR) with high forward skewness (+1.74) and phase sensitivity to the magnetic field. Overall, our findings suggest that the Nb—1D $Bi_4Br_4$—Nb weak link is a promising platform for on-chip superconducting magnetic sensors, compatible with scalable nanofabrication and broader development of topological-superconductor hybrid electronics.

## Introduction

When two superconductors are connected by a "weak link," they form a device that exhibits the Josephson effect, enabling a dissipationless supercurrent to flow between them. This supercurrent is characterized by a unique current–phase relationship (CPR), which can be used in magnetic sensing and superconducting interferometry [1-4]. Weak link materials are usually quasi-one-dimensional (1D) elements and can be metallic, semiconducting, (super)conducting nanobridge constrictions, or tunnel barriers (insulators), with transport properties ranging from strongly diffusive to nearly ballistic. The full description of the supercurrent involves the Andreev process, in which electrons convert into holes at superconductor–weak link interfaces. This process can generate phase-coherent electronic bound states across the weak link. According to established quantum transport models, the "weak link" transparency and effective length determine the critical current, the shape of the CPR, and its response to magnetic fields [4,5].

In particular, high transmission coefficients increase the maximum Josephson current per mode and effectively reduce current dissipation within the junction when superconducting [5,6]. This significantly improves phase sensitivity. All these unique features of the junction are directly useful in magnetometry. In superconducting magnetometry, a superconducting phase difference across a weak link causes a change in magnetic flux, generating a signal for magnetic flux transduction. This flux variation also results in oscillations in the voltage depending on the applied field or critical current. Superconducting quantum interference devices (SQUIDs) and related nanoSQUIDs operate on this principle [6] to ensure stable operation across a broad spectrum of applied magnetic fields. As a result, Josephson weak link applications have become highly attractive at the nanoscale. Another reason for these promising applications is that weak links can operate ultra-fast and with low power in switching device applications [6,7]. These devices can also function under higher magnetic fields and critical current densities than traditional junction devices, while still providing a strong response to magnetic flux changes in an interference loop application [6,7].

Alongside advanced Josephson junction electronics, two-dimensional (2D) topological insulators have recently become more prominent. They block current flow through their bulk but support unidirectional current via topologically protected boundary states—edge states [8-11]. Electron transport through these edge states is generally immune to backscattering because they are protected by topological features like spin-momentum locking caused by specific symmetries and the spin-orbit coupling effect [8-11]. Additionally, these edge states exhibit the quantum spin Hall effect (QSH) and quantized spin Hall conductivity. In these quantum insulators, topological electrons can move through two counterpropagating edge states with opposite spins. Unlike trivial nanowires, topological edge states can persist even in the presence of impurities or disorder [12-14]. This makes topological edge states attractive because they lessen the need for high upper-critical magnetic field superconductors in certain applications. Their unique properties have generated interest in various superconducting devices, especially those involving proximity effects. Intriguingly, superconductivity induced via the proximity effect can coexist with topological edge states when a conventional superconductor contacts a 2D topological insulator [15-17]. This kind of superconductivity opens new options for designing phase-coherent Josephson devices. Recent experiments have reported, notably using HgTe—a 2D topological insulator—as a weak link between superconductors [18,19]. These studies focused on Josephson coupling and transport dominated by edge states [18,19]. However, maintaining proper parity relaxation and fermion parity remains quite challenging in a 2D HgTe-based junction [20-26], and since topological edge states exist only at very low temperatures (around 10-25 K), this poses a significant obstacle to practical applications [27,28].

Unlike 2D HgTe, $\alpha$-$Bi_4Br_4$ forms a stable layered structure at room temperature [29-31], where two superconducting phases have been observed: one with a $T_c$ of 6.8 K that appears under 3.8 GPa, and another with a $T_c$ of 9 K that emerges under 5.5 GPa [32]. The second superconducting phase results from a transition from monoclinic (C2/m) to triclinic ($P\bar{1}$) structure, while the first phase can occur without any structural change, indicating the coexistence of the initial and topological phases [32]. Monolayer α-$Bi_4Br_4$ (hereafter called 2D $Bi_4Br_4$) is a 2D topological material with a notable insulating surface bandgap ($E_g$>0.18 eV) [29], effectively preventing bulk surface quasiparticles from entering the system. This facilitates a highly resilient parity relaxation process in the robust edge states of 2D $Bi_4Br_4$. Recently, QSH edge states have been observed at room temperature in 2D $Bi_4Br_4$ [33]. Additionally, the surface of 2D $Bi_4Br_4$ lacks dangling bonds, enabling cleaner interface formation with superconductors. These distinct properties of 2D $Bi_4Br_4$ strongly motivate further research into weak links that use the topological edge states of 2D $Bi_4Br_4$ as the main pathway for Josephson current.

In this study, we computationally analyze a superconducting Josephson junction using bulk Nb as the superconducting leads and topological edge states of 2D $Bi_4Br_4$ as the link, creating a Nb-1D $Bi_4Br_4$–Nb Josephson device. We observe nearly ballistic transmission through the weak link, placing the junction in a regime where Andreev bound-state (ABS) transport is efficient and highly phase-coherent. Additionally, we find that the device is multichannel at the microscopic level. The CPR relation exhibits non-sinusoidal behavior with highly skewed current and phase sensitivity, emphasizing its potential for magnetic field sensing. This work offers a practical pathway for designing topology-enabled, application-ready nanoscale magnetometers.

**Methods**

We perform bulk and slab electronic-structure calculations within density-functional theory (DFT) using the plane-wave (PW) pseudopotential method as implemented in Quantum ESPRESSO (QE) [34-36]. We utilize the generalized-gradient approximation (GGA) in the Perdew–Burke–Ernzerhof (PBE) form to treat exchange–correlation effects [37,38]. In all calculations, we explicitly include the spin–orbit coupling (SOC) effect in every self-consistent calculation. We set 65 Ry as the plane wave cutoff energy and specify $10^{-10}$ Ry and $10^{-7}$ Ry/Au for total energy and force convergence, respectively. For bulk, slab, and edge calculations, we use $16 \times 16 \times 16$, $16 \times 16 \times 1$, and $16 \times 1 \times 1$ k-points in the Brillouin zone integration. For the slab and nanoribbon calculations, we employ the Engel-Vosko (EV93) functional to improve band ordering and edge-state placement in reduced-dimensional geometries. We also perform electron-phonon coupling calculations using density functional perturbation theory (DFPT) for bulk, surface, and edge in QE. In electron-phonon calculations, we use 888, 881, and 811 q-points for bulk, slab, and edge, respectively.

To obtain a compact real-space Hamiltonian for mesoscopic transport, we construct maximally localized Wannier functions (MLWFs) and the corresponding tight-binding (TB) Hamiltonian using Wannier90 [39-42]. Specifically, we express the Hamiltonian as [39-42]

$$H = \sum_{\mathbf{R}} \sum_{i,j} c_i^{\dagger}(\mathbf{0})\, H_{ij}(\mathbf{R})\, c_j(\mathbf{R}), \qquad (1)$$

where $c_i^{\dagger}(\mathbf{R})$ builds an electron in a Wannier orbital $i$, which is located at the position in the lattice $\mathbf{R}$, and $H_{ij}(\mathbf{R})$ stand for the hopping and onsite matrix elements. In this Wannier basis, we can generally express the real-space Hamiltonian matrix elements as [39-42]

$$H_{mn}(\mathbf{R}) = \langle w_{m\mathbf{0}} \mid \hat{H} \mid w_{n\mathbf{R}} \rangle = \frac{1}{N_k} \sum_{\mathbf{k}} e^{-i\mathbf{k}\cdot\mathbf{R}} \left[U^{\dagger}(\mathbf{k})\, E_n(\mathbf{k})\, U(\mathbf{k})\right]_{mn}, \qquad (2)$$

where $E_n(\mathbf{k})$ represents the diagonal matrix of selected Kohn–Sham eigenvalues ($m$ and $n$ are the band indices) at momentum $\mathbf{k}$, and $U(\mathbf{k})$ stands for the unitary rotation used to define the Wannier gauge. We use this Hamiltonian to obtain bulk electronic dispersions. For edge (1D) transport along a fixed direction $\hat{\mathbf{d}}$, we consider $\mathbf{k} = k\,\hat{\mathbf{d}}$, based on which we compute edge electronic dispersion. Then, we construct the Bloch Hamiltonian as [39-42]

$$H(\mathbf{k}) = \sum_{\mathbf{R}} e^{i2\pi\mathbf{k}\cdot\mathbf{R}}\, H(\mathbf{R}). \qquad (3)$$

We finally generate the TB Hamiltonian for bulk Nb, the $Bi_4Br_4$ slab, and the $Bi_4Br_4$ nanoribbon. The k-point meshes used in Wannierization are $16 \times 16 \times 16$, $16 \times 16 \times 1$, and $16 \times 1 \times 1$ for bulk, slab, and edge, respectively, ensuring they are consistent with the underlying DFT sampling.

We obtain a surface model from the bulk Wannier Hamiltonian by truncating along the surface-normal direction and treating the remaining directions as periodic. We construct nanoribbons through finite-size truncation in two directions ($y$ and $z$) with periodicity along the ribbon axis ($x$). In device simulations, we rescale the nanoribbon Hamiltonian to ensure the ribbon is sufficiently large (36-250 nm) to minimize finite-size confinement effects. Then, we compute the self-energies of the semi-infinite leads (the bulk Nb-nanoribbon interface) using an iterative surface Green's function approach [43]. We model the hybrid interface at the TB level by coupling a TB Hamiltonian to the topological edge—that is, a TB Hamiltonian of Nb ($H_{\mathrm{Nb}}$) to the edge Hamiltonian $H_{\mathrm{edge}}$. Using an interface hopping matrix $V$, we construct the device Hamiltonian (H) in the Wannier basis as

$$H_{\mathrm{device}} = H_{\mathrm{edge}} + H_{\mathrm{Nb}} + H_{\mathrm{int}}, \qquad (4)$$

where the interface Hamiltonian is defined by

$$H_{\mathrm{int}} = \sum_{i \in \mathrm{edge}} \sum_{j \in \mathrm{Nb}} \left(c_i^{\dagger} V_{ij} c_j + \mathrm{h.c.}\right).$$

Here, $i$ and $j$ represent localized Wannier orbitals near the interface on the two Nb sides. We construct $V_{ij}$ by using the dominant interface orbitals derived from spatial localization and Wannier projections. Then, we assign symmetry-allowed couplings to ensure consistency with the orbital characters and the relative configuration. To model a finite-length device, we repeat the Wannier Hamiltonian of the unit cell along the transport direction to build a nanoribbon Hamiltonian of length $L = Na$, where $a$ stands for the lattice constant of the ribbon and $N$ represents the desired number of repeated unit cells. We then write the large nanoribbon Hamiltonian as,

$$H_{\text{big}} = \begin{pmatrix} H_0 & H_1 & H_2 & \cdots & H_{N-1} \\ H_1^\dagger & H_0 & H_1 & \cdots & H_{N-2} \\ H_2^\dagger & H_1^\dagger & H_0 & \cdots & H_{N-3} \\ \vdots & \vdots & \vdots & \ddots & \vdots \\ H_{N-1}^\dagger & H_{N-2}^\dagger & H_{N-3}^\dagger & \cdots & H_0 \end{pmatrix} \quad (5),$$

where each block $H_t$ represents hopping between repeated unit cells separated by lattice spacings along the periodic direction of the $Bi_4Br_4$ nanoribbon. We construct this Hamiltonian by assembling Wannier hopping matrices based on open boundary conditions. Then, we can write the Fermi crossing for a given chemical potential ($\mu$) as $E_n(k_{F,n}) = \mu$, and we extract the corresponding Fermi velocity ($v_{F,n}$) from the electronic band dispersion of the large nanoribbon, where $v_{F,n} = \frac{1}{\hbar}\frac{\partial E_n(k)}{\partial k}|_{k=k_{F,n}}$. We define the coherence length for each crossing (mode) using an induced superconducting gap as

$$\xi_n = \frac{\hbar \mid v_{F,n} \mid}{\Delta}. \quad (6)$$

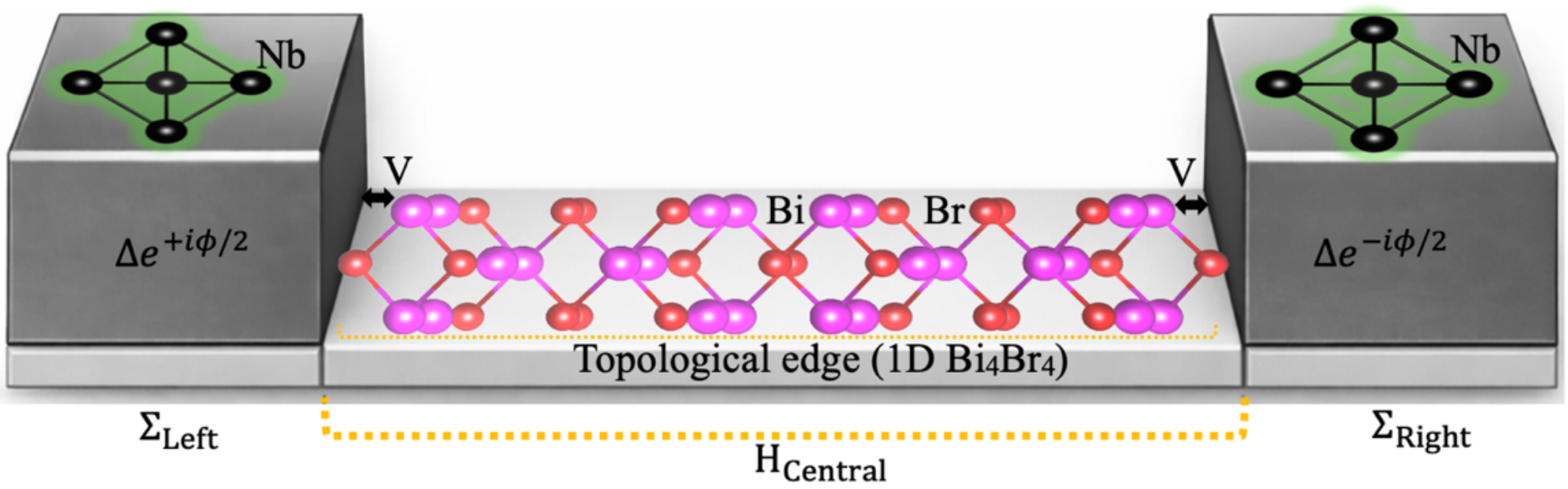

**Figure 1. Schematic illustration of the Nb-1D $Bi_4Br_4$–Nb weak-link device and the approach used for modeling it.**

Schematic structure of the Nb-1D $Bi_4Br_4$–Nb Josephson junction. A finite topological insulator, $Bi_4Br_4$ nanoribbon, connects two superconducting Nb leads with a phase difference. The topological nanoribbon hosts metallic edge channels that serve as weak links between the superconducting Nb leads. The finite-length nanoribbon (central region) is coupled to left and right superconducting Nb leads through interface hopping matrices. The Hamiltonian is built in a Wannier basis and analyzed with the Bogoliubov–de Gennes (BdG) formalism. Lead self-energies account for the semi-infinite superconducting electrodes (bulk Nb).

Now, for simplicity, we introduce a dimensionless parameter ($\lambda$) that characterizes the quasiparticles' phase accumulation in the weak link, where

$$\lambda_n = \frac{L}{\xi_n} = \frac{L\Delta}{\hbar \mid v_{F,n} \mid}. \quad (7)$$

Here, channel (mode) manifolds of the same $\lambda$ share the same ABS dispersion. In the absence of backscattering, we can state the condition for ballistic quantization in the ABS spectrum of a two-mode system as [44]

$$\sin\left(\epsilon\lambda_1 - s\frac{\phi}{2} - \arccos\epsilon\right)\ \sin\left(\epsilon\lambda_2 + s\frac{\phi}{2} - \arccos\epsilon\right) = 0, \quad (8)$$

where $\epsilon = E_A/\Delta$ and $\phi$ represent the normalized ABS energy and the superconducting phase difference, respectively. Here $s = \pm 1$ stands for the two time-reversed modes. This equation yields the ballistic spectrum of two distinguishable ABS related to $\lambda_1$ and $\lambda_2$. The effect of backscattering in the weak link can be modeled by an effective scatterer characterized by a transmission probability $\tau$ and a scatterer location $x_r$ within the junction. We then write the ABS energies in a transcendental equation of the form [44]

$$\tau\cos\left[\frac{1}{2}(\lambda_1 - \lambda_2)\epsilon \mp \phi\right] + (1-\tau)\cos\left[\frac{1}{2}(\lambda_1 + \lambda_2)\epsilon + x_r\right] \\ = \cos[2\arccos\epsilon - (\lambda_1 + \lambda_2)\epsilon], \quad (9)$$

whose limit is equal to **eq. 8** in the ballistic case ( $\tau \to 1$). In a microscopic model, we can superimpose all solutions of the corresponding parameter sets $(\lambda_{1,i}, \lambda_{2,i}, \tau_i, x_{r,i})$ to capture multichannel transport in the device. For modeling this device (as shown in **Fig. 1**), i.e., superconductor (Nb)—topological edge states—Nb (STS) junction, we use the pairing potential as a diagonal matrix in the Wannier basis, i.e.

$$\Delta(x) = \begin{cases} \Delta_{\mathrm{SC}}\, e^{+i\frac{\phi}{2}}, & x \in \text{left superconductor (Nb)}, \\ 0, & x \in \text{normal region (topological edge states)}, \\ \Delta_{\mathrm{SC}}\, e^{-i\frac{\phi}{2}}, & x \in \text{right superconductor (Nb)}. \end{cases} \quad (10)$$

Eventually, we write the full Bogoliubov-de Gennes (BdG) Hamiltonian as [45-47]

$$H_{\mathrm{BdG}}(\phi) = \begin{pmatrix} H_{\mathrm{big}} - \mu I & \Delta(\phi) \\ \Delta^{\dagger}(\phi) & -\left(H_{\mathrm{big}} - \mu I\right)^{T} \end{pmatrix}. \quad (11)$$

We obtain its associated eigenvalues $E_n(\phi)$ and eigenvectors that describe electron and hole amplitudes. We identify the subgap eigenstates from this equation (which satisfy $0 < E_n < \Delta$) as Andreev bound states. Then, we calculate the quasiparticle density of states (DOS) in the superconducting state by computing the retarded Green's function $G^r(E)$ as.

$$\rho(E) = -\frac{1}{\pi}\,\mathrm{Im\,Tr}\, G^r(E), \quad (12)$$

We now describe the superconducting local density of states (LDOS) at an orbital $i$ as

$$\rho_i(E,\phi) = \sum_m [\mid u_m(i,\phi) \mid^2 \ \delta(E - E_m) + \mid v_m(i,\phi) \mid^2 \ \delta(E + E_m)].$$

Here, $u_m$ and $v_m$ represent the electron and hole components of the BdG eigenvector. We also calculate the Bardeen-Cooper-Schrieffer (BCS) density of states (DOS) for a uniform $s$-wave gap $\Delta$ based on, where

$$N_s(E) = N_0 \,\mathrm{Re}\left[\frac{\mid E \mid}{\sqrt{E^2 - \Delta^2}}\right]. \qquad (13)$$

To evaluate transport properties, we use the non-equilibrium Green's function (NEGF) formalism, where we model the scattering region as a nanoribbon, and the semi-infinite superconducting leads are included via self-energies. For an open system, we can write the retarded Green's function as

$$G^r(E) = [(E + i\eta)I - H_{\mathrm{BdG}} - \Sigma_L(E) - \Sigma_R(E)]^{-1}. \qquad (14)$$

Based on this, we calculate the lead-induced broadening matrices using this equation.

$$\Gamma_\alpha(E) = i\left[\Sigma_\alpha(E) - \Sigma_\alpha^\dagger(E)\right], \alpha \in \{L, R\}, \qquad (15)$$

and the total transmission calculated using the Caroli formula [48], which can be expressed as

$$T(E) = \mathrm{Tr}[\Gamma_L(E)\, G^r(E)\, \Gamma_R(E)\, G^a(E)], \qquad (16)$$

with $G^a = (G^r)^\dagger$. Now we can also calculate the normal state (two-terminal) linear conductance using the Landauer relation [49,50].

$$G_N = \frac{e^2}{h} T(E_F). \qquad (17)$$

Finally, we determine the superconducting differential conductance by evaluating the electron–hole (Andreev) and transmission probabilities in Nambu space. For energy $E$, we express the zero-temperature two-terminal differential conductance as [51]

$$G_S(E) = \frac{2e^2}{h}[N(E) - R_{ee}(E) + R_{he}(E)]. \qquad (18)$$

Here $R_{ee}$ and $R_{he}$ capture the Andreev reflection for this two-terminal differential conductance. At equilibrium, we obtain the Josephson current by calculating the phase dependence of the ABS eigenvalues, which is represented as [52]:

$$I(\phi) = \frac{2e}{\hbar} \sum_{m>0} \frac{\partial E_m(\phi)}{\partial \phi} \tanh\left(\frac{E_m(\phi)}{2k_{\mathrm{B}}T}\right). \qquad (19)$$

## Results & discussion

The calculated bulk band structure (**Fig. 2a**) of Nb shows nearly free-electron-like dispersion, with multiple bands crossing the Fermi level. These metallic states form a high-density pool of delocalized states. This electronic structure supports conventional phonon-mediated superconductivity in bulk Nb (see below). As a result, the system is ideal for inducing superconductivity across a weak link via the proximity effect. Conversely, the bulk 2D surface of $Bi_4Br_4$ exhibits a clear gap (~0.18 eV) at the zone center (**Fig. 2b**). This calculated band gap closely matches experimental measurements [29,30]. Compared to these gapped surface states, the electronic states of the 1D nanoribbon feature gapless metallic states at the Fermi level (**Fig. 2c**). Notably, the ribbon edge states are considered to be in a normal (non-superconducting) state in the isolated ribbon calculation. In this normal state, there is no intrinsic pairing instability within the topological edge states. Consequently, these edge states act as normal metallic channels with topological protection against backscattering. They only become superconducting through the proximity effect from Nb in the device setup. Therefore, throughout this work, the Josephson junction device is described as S (Nb) – N (topological edge states) – S (Nb) (**Fig. 1**).

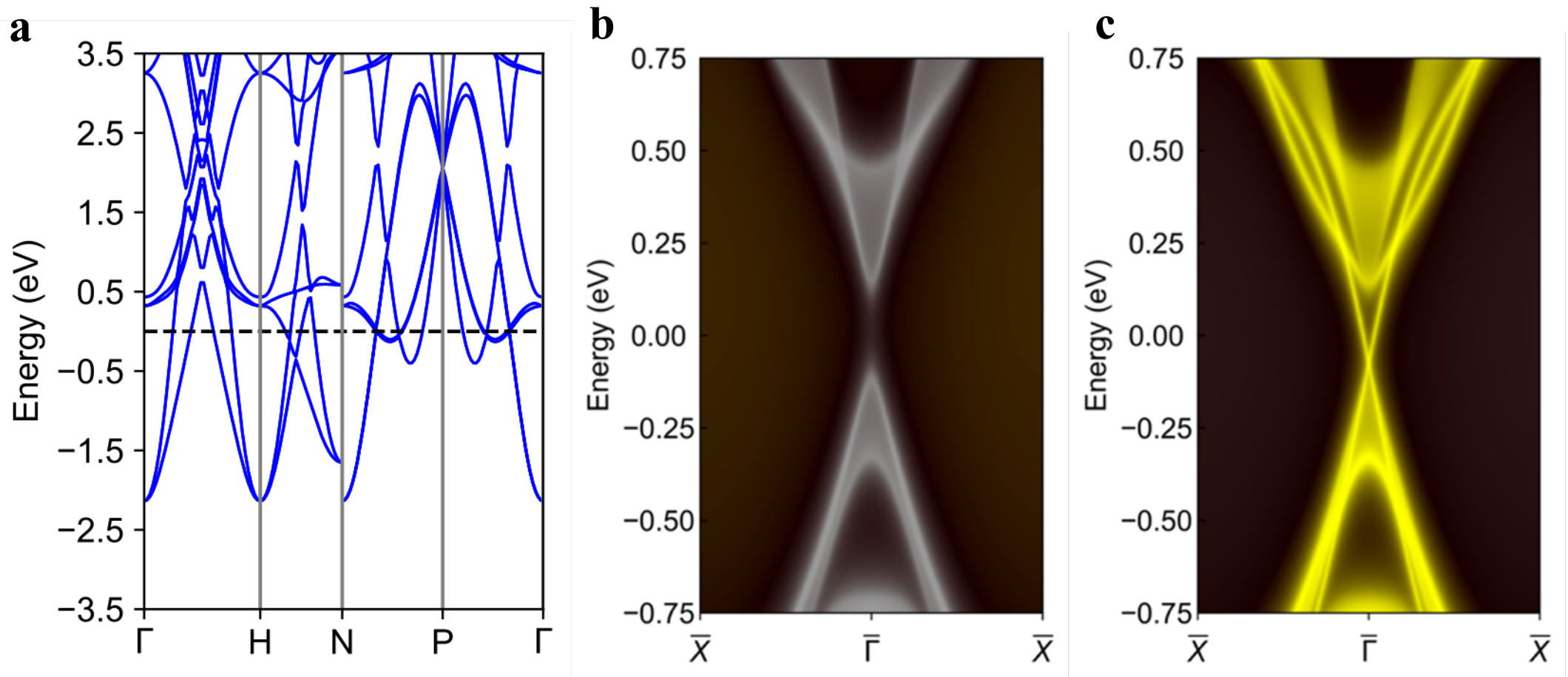

**Figure 2. Bulk electronic band structure of the constituent materials Nb and the one-dimensional electronic dispersion of $Bi_4Br_4$.**

(**a**) Bulk electronic band structure of superconducting Nb calculated using DFT with spin–orbit coupling. (**b**) Electronic band spectrum of two-dimensional bulk surface $Bi_4Br_4$. We use a 1D BZ in this plot for comparison. (**c**) Electronic dispersion of the topological edge (nanoribbon) including SOC effects. Two topologically protected, counterpropagating channels cross the Fermi level (indicated by zero energy).

In the bulk Nb unit cell, two Nb atoms give rise to three acoustic and three optical phonon branches. The phonon spectrum of these six modes shows positive (real) frequencies across the entire Brillouin zone (BZ) (**Fig. 3a**). These positive frequencies signify a dynamically stable Nb lattice and allow for standard electron–phonon pairing analysis within the Migdal–Eliashberg framework. We calculate the Eliashberg spectral function $\alpha^2F(\omega)$ for phonon frequency $\omega$ using the equation [39,42,53].

$$\alpha^2F(\omega) = \frac{1}{N(E_F)} \sum_{\mathbf{k},\mathbf{q},\nu} |\, g^{(\nu)}_{\mathbf{k},\mathbf{k+q}} \,|^2 \;\; \delta(\epsilon_{\mathbf{k}})\, \delta\big(\epsilon_{\mathbf{k+q}}\big)\, \delta\big(\omega - \omega_{\mathbf{q}\nu}\big). \qquad (20)$$

Here $N(E_F)$ represents the density of states at the Fermi level, and $g_{\mathbf{k},\mathbf{k+q}}^{(\nu)}$ are electron–phonon matrix elements for the phonon mode $\nu$ with momentum $\mathbf{q}$. The spectrum indicates that low-frequency phonon modes are strongly coupled to electrons, resulting in a significant contribution to the electron-phonon coupling $\lambda$. To quantify this, we calculate the electron–phonon coupling strength using $\alpha^2F(\omega)$, which is defined as [39,53]

$$\lambda = 2\int_0^\infty d\omega \frac{\alpha^2F(\omega)}{\omega}. \qquad (21)$$

The electron-phonon coupling constant $\lambda$, with a value of 1.19, indicates a strong-coupling regime. Based on the electron-phonon coupling strength and the Eliashberg spectral function, we calculate the logarithmic-average phonon frequency

$$\omega_{\log} = \exp\left[\frac{2}{\lambda}\int_0^\infty d\,\omega \frac{\alpha^2F(\omega)}{\omega}\ln\,(\omega)\right]. \qquad (22)$$

Together with a physically reasonable Coulomb pseudopotential $\mu^* = 0.1$, we calculate the superconducting transition temperature using the McMillan equation given by [54]

$$T_c = \frac{\omega_{\log}}{1.2}\,\exp\left[-\frac{1.04(1+\lambda)}{\lambda-\mu^*(1+0.62\lambda)}\right]. \qquad (23)$$

Using this equation for bulk Nb yields a $T_c\ {\sim}9\,K$ that is very close to the known bulk value $9.2 - 9.5\,K$. Thus, this superconducting description of the leads falls within the expected weak-coupling regime relevant for Josephson junctions. We therefore treat the superconducting electrodes (Nb) as a conventional BCS-like s-wave superconductor with a calculated gap $\Delta(T) \cong 4$ meV. Such a unique feature of Nb makes weak-link physics and Josephson transport analysis well-grounded.

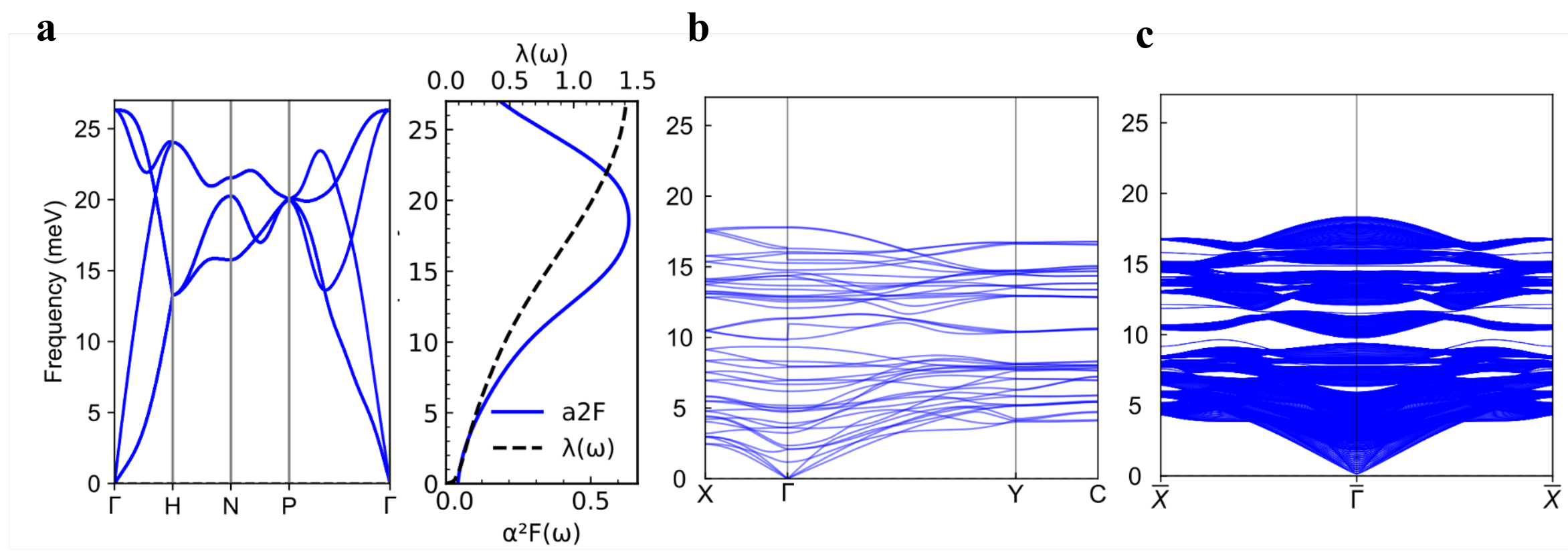

**Figure 3. Bulk phononic band structure of constituent materials Nb, and 1D phononic dispersion of $Bi_4Br_4$.**

**(a)** Bulk phononic band structure, Eliashberg spectral function, and cumulative electron-phonon coupling constant ($\lambda$) of superconducting Nb calculated within DFPT including spin–orbit coupling (SOC) effect. **(b)** Phononic band spectrum of 2D bulk surface $Bi_4Br_4$. (**c**) Phononic dispersion of the topological (1D) nanoribbon including the SOC effect.

Compared to superconducting Nb electrodes, the bulk surface of 2D $Bi_4Br_4$ has 16 atoms per unit cell, resulting in 3 acoustic and 45 optical phonon branches (**Fig. 3b**). All these phonon modes show positive (real) frequencies, indicating that 2D $Bi_4Br_4$ is also dynamically stable. Since phonon calculations are computationally very intensive for a mesoscopic nanoribbon, we use a small-sized nanoribbon to compute its phonon dispersion. The phonon dispersion of this topological nanoribbon similarly demonstrates dynamical stability (**Fig. 3c**). While the ribbon exhibits well-defined vibrational branches, including low-frequency modes due to reduced dimensionality, these phonon modes and the metallic electronic nature of the nanoribbon alone do not imply superconductivity at the edge. In fact, Haque *et al.* showed that superconductivity in topological edge states is absent due to topological protection [55]. Because we used the same nanoribbon geometry as in Ref. [55], we can confidently treat the edge states as normal metallic states with topological protection against backscattering.

These topological edge states connect two superconducting electrodes (Nb) with a phase difference ($\delta = \pm\phi/2$) (**Fig. 1**), where the edge functions as a finite-length normal (edge) region of length $L$ within a Josephson weak link. Throughout this normal region, quasiparticle transport is governed by Andreev reflections at the superconductor–normal (SN) interfaces, where these reflections are repeated. As a result, an incident electron is retroreflected from the left SN interface as a hole, and a Cooper pair is added to the superconducting condensate of the left electrode (S). This reflected hole from the left SN propagates through the normal region (topological edge states) and approaches the right SN, where the hole is also retroreflected as an electron, and a Cooper pair is moved out from the right superconducting condensate. This process transfers one Cooper pair from the right to the left superconducting electrode. This coherent electron–hole cycle induces a quantized Andreev bound state (ABS) within the superconducting gap. This ABS carries the Josephson supercurrent [56].

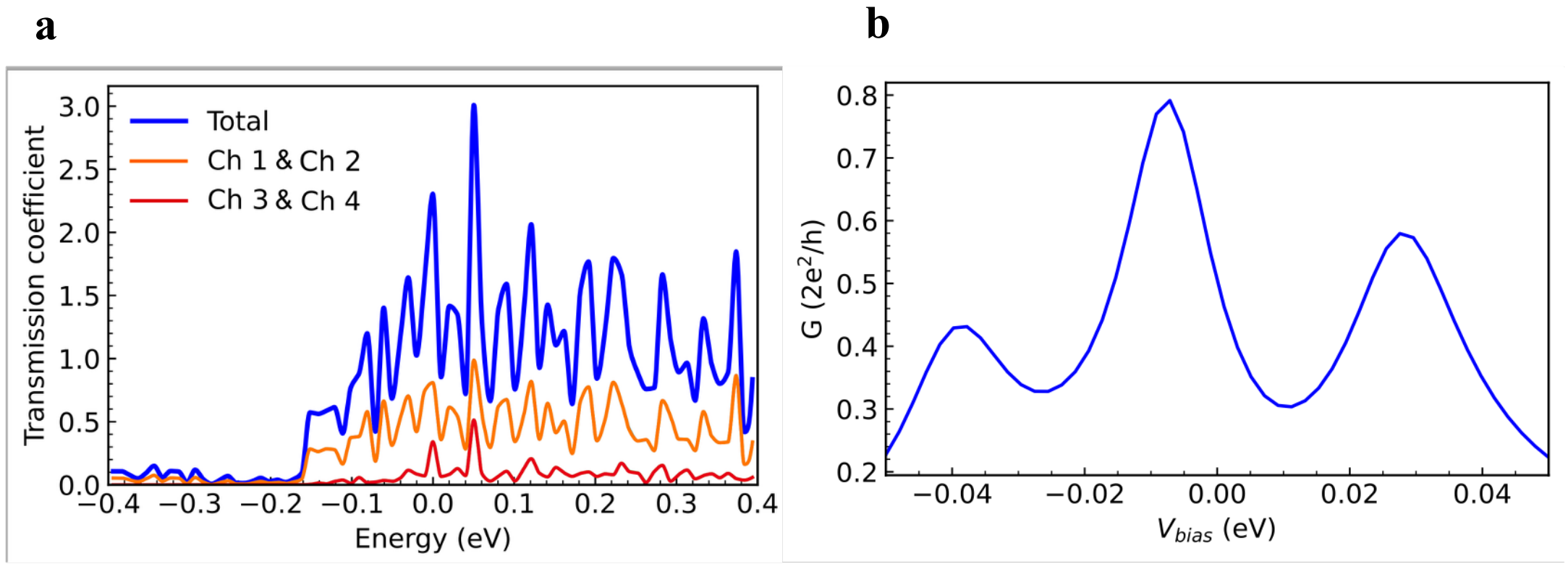

**Figure 4. Calculated normal-state transmission and superconducting subgap conductance of Nb—1D $Bi_4Br_4$—Nb weak link.**

(**a**) Energy-dependent total transmission coefficient $T(E)$ decomposed into individual transmission eigenchannels $\tau_n$. (**b**) Superconducting differential conductance $G_{\text{tot}}(V)$ (normalized to $2e^2/h$) calculated within the Bogoliubov–de Gennes and non-equilibrium Green's-function formalism. The total conductance emphasizes the dominance of Andreev processes in the subgap regime.

The phase-dispersive nature of ABS critically depends on achieving high transparency across multiple channels. Since the topological edge remains normal when isolated, we now quantify this through channel-resolved transmission calculations. We evaluate the transparency of the Nb—1D $Bi_4Br_4$—Nb weak link in both the normal and superconducting states (**Fig. 4**), demonstrating that the weak-link device operates in a high-transmission, few-mode regime. We observe two channels transmitting strongly within the relevant low-energy window, while the third and fourth channels remain partially open. Additionally, a fifth and remaining channel is effectively closed. This is expected because the weak link is formed by the topological edge of 1D $Bi_4Br_4$, which hosts two counterpropagating helical (topological) edge modes that connect the superconducting electrodes (Nb). Accounting for spin degeneracy, this yields four transport channels in total, with two left-moving and two right-moving edge modes contributing to the ABS. Therefore, transport in Nb—1D $Bi_4Br_4$—Nb is dominated by a small number of highly transparent modes, consistent with edge-dominated conduction in narrow 1D $Bi_4Br_4$. Notably, the high transmission in two channels indicates that the Nb—1D $Bi_4Br_4$ interface and the finite junction region cause only minimal normal backscattering for these leading channels. This can be attributed to the topological protection of the normal region (1D $Bi_4Br_4$), which safeguards against backscattering.

From the Landauer perspective, these channels are expected to control the normal-state two-terminal conductance near the Fermi level (see **eq. 17**). To demonstrate this, we calculate the corresponding superconducting differential conductance using the Nambu-space NEGF. In this calculation, we treat the finite device region with a BdG Hamiltonian and apply pairing to the left and right superconducting electrodes (lead portions) of the device. The leads are considered semi-infinite and included as embedding self-energies in Nambu space. We evaluate the total subgap conductance $G_{\text{tot}}$ at each bias point $V$ (where $V \in [-0.02, 0.02]$ eV with $\phi = 0$). Open channels with strong coupling to proximity-induced superconducting states dominate. As a result, high normal-state transparency leads to efficient Andreev reflection and higher subgap conductance. The calculated conductance $G_{\mathrm{tot}}(V)$ directly reflects how much the low-bias response originates from Andreev conversion. This also provides a quantitative link between the high-$\tau_n$ channels and the subgap conductance features typically used to diagnose highly transparent weak links [57]. High conductance values also suggest the presence of strongly phase-dispersive Andreev bound states in the weak link device.

To visualize such an ABS of Nb-1D $Bi_4Br_4$-Nb weak link, we systematically start from a single channel at the short-junction limit. In this limit, the ABS energy dispersion is governed by $E_A(\phi) = \Delta\sqrt{1 - \tau \sin^2(\phi/2)}$ where $\Delta$ represents the superconducting gap. Using the calculated transmission coefficient ($\tau = 0.86$), the ABS spectra of a single channel exhibit a highly dispersive nature (**Fig. 5a**). Here, a single coherent transport channel connects the two superconducting (Nb) electrodes, and the resulting ABS spectrum depends solely on the superconducting phase difference ($\phi$) and the normal-state transmission coefficient ($\tau$) of that channel. Consequently, a high transmission coefficient is associated with the dispersive behavior of ABS. Such high dispersion can lead to a larger supercurrent contribution per channel. When the weak link becomes perfectly transparent ($\tau \rightarrow 1$), the ABS becomes maximally phase-sensitive, reflecting nearly ballistic transport through the weak link [52].

In a practical scenario, the Nb-1D $Bi_4Br_4$-Nb weak link features multiple channels. We therefore extend the previous approach to the multichannel regime (**Fig. 5b**), which is a physically relevant setting for the Nb-1D $Bi_4Br_4$-Nb weak link. When several edge-derived channels participate in quantum transport, each one contributes to its own ABS. As a result, these channels give rise to multiple ABS modes within the superconducting gap of the electrodes. Strong spin–orbit coupling (SOC) in 1D $Bi_4Br_4$ and the finite length of the junction cause these branches to have different Fermi velocities and transmission coefficients, leading to distinct phase dispersions. The strong SOC effect produces spin-split**,** or helical, dispersions, visualized as two shifted parabolas in 1D (**Fig. 5b**). Therefore, right- and left-moving Nb states can acquire different spin textures. These states provide a platform for multichannel Josephson physics, in which superconducting correlations induced by Nb act on the SOC-modified spectrum of 1D $Bi_4Br_4$. These interactions generate ABS (**Fig. 5a-b**), whose phase dispersion depends heavily on the transmission coefficient of each participating channel.

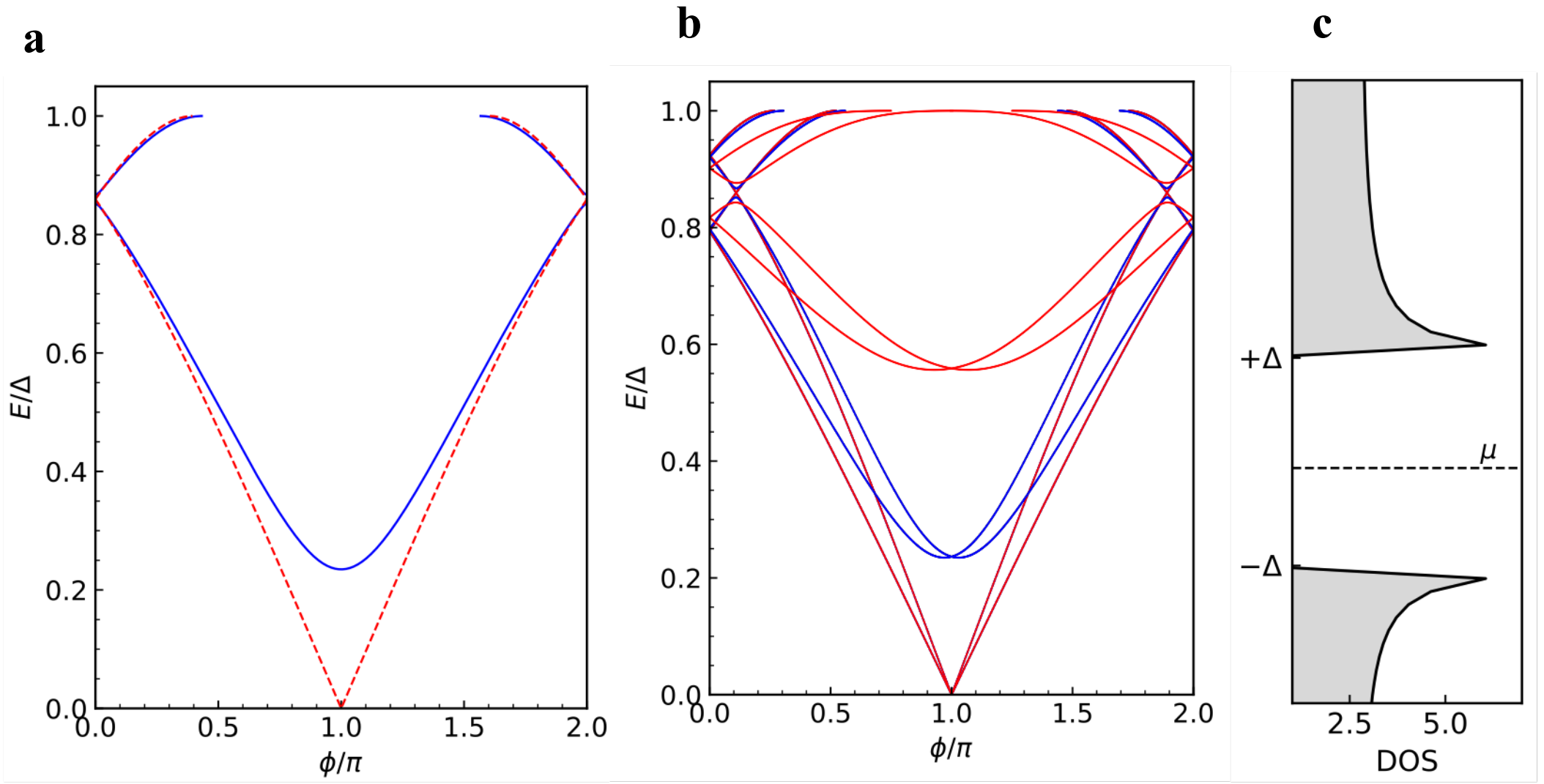

**Figure 5. Andreev bound states of Nb—1D $Bi_4Br_4$—Nb** weak link**, considering both single-channel and multichannel models.**

(**a**) Andreev bound-state spectrum in the single-channel limit, where a single coherent transport channel produces a pair of phase-dependent subgap states whose dispersion is determined by the channel transmission ($\tau$). (**b**) Andreev bound-state spectrum in the multichannel regime, where multiple conducting modes contribute distinct subgap branches; interchannel coupling and normal backscattering lead to avoided crossings, channel-dependent phase dispersions (**c**) Local superconducting density of states (DOS) labelling the superconducting gap ($\Delta$) and the Fermi level ($\mu$).

Furthermore, any normal backscattering or intermode coupling can mix two channels, leading to avoided crossings in the ABS spectrum of Nb—1D $Bi_4Br_4$—Nb (**Fig. 5b**), at points where isolated ABS modes would normally intersect. In an idealized transparent weak link, these subgap modes could cross as a function of phase. However, with any normal backscattering, where reflection is finite, i.e., $\tau < 1$, two channels can hybridize counterpropagating amplitudes. Physically, these avoided crossings in the spectra of Nb-1D $Bi_4Br_4$-Nb serve as a spectral fingerprint of imperfect mode matching. They also reveal intermode coupling within the weak link and influence how the ABS spectrum responds to external perturbations. As a result, the supercurrent becomes sensitive to disturbances such as disorder, electrostatic gating, and magnetic fields. These avoided crossings provide information about channel-dependent transmission and scattering processes and are essential in shaping the Josephson current–phase relation (CPR). This single-channel versus multichannel distinction forms the basis for the subsequent analysis of the Nb—1D $Bi_4Br_4$—Nb weak-link device, especially regarding the ABS spectrum and CPR.

We calculate the ABS spectrum as a function of the superconducting phase difference ($\phi$) and for four different ribbon thicknesses, derived from the full BdG Hamiltonian of the proximitized Nb—1D $Bi_4Br_4$—Nb weak link. A key feature of this BdG spectrum for Nb—1D $Bi_4Br_4$—Nb is its inherent particle–hole symmetry. This symmetry guarantees that for every ABS at a given energy $+E_n(\phi)$, there is a corresponding ABS at the mirrored energy $-E_n(\phi)$. As a result, only the branches with positive energy are physically independent. This symmetry also ensures that the Josephson current calculated from the spectrum is an odd function of phase, i.e. $I(-\phi) = -I(\phi)$, only if other symmetries, such as inversion symmetry, are not broken. The dependence of ABS on thickness (**Fig. 6**) arises from quantum confinement effects and the reconstruction of subgap modes. The reconstruction arises from ABS wavefunction hybridization due to thickness variations, reshaping the phase-dispersive subgap ABS spectrum. Specifically, changing the ribbon thickness affects the number of transverse modes, their Fermi velocities, and their electronic character, including orbital and spin attributes. Since the Nb—1D $Bi_4Br_4$—Nb weak link is finite along the transport direction, both the superconducting gap and a traversal (Thouless-like) scale ($E_{\mathrm{Th}} \sim \hbar v_F/L$) define the characteristic ABS energy range. Therefore, adjusting the structural parameters of 1D $Bi_4Br_4$, such as its thickness, significantly alters the ABS dispersion. Such modifications also shift avoided crossings, alter the minigap, and change the density of low-energy states. These low-energy states primarily influence the phase-coherent transport of the Nb—1D $Bi_4Br_4$—Nb weak link.

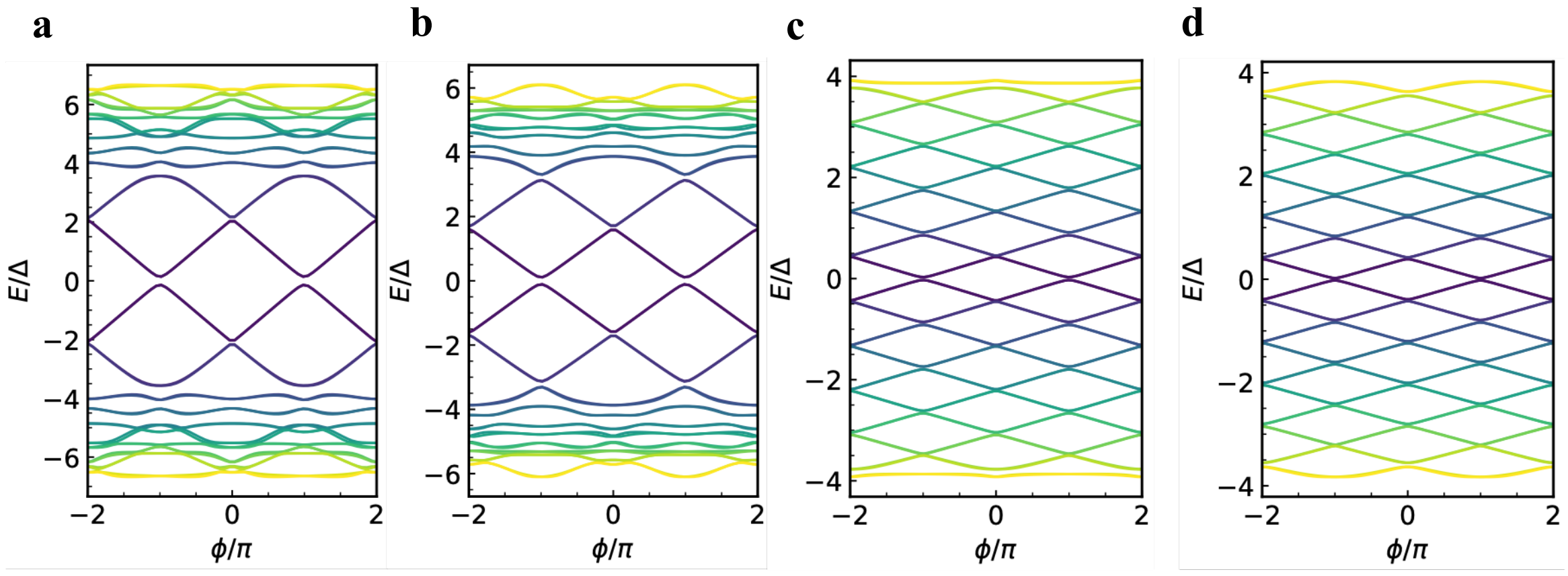

**Figure 6. Calculated Andreev bound states in Nb—1D $Bi_4Br_4$—Nb weak link from full BdG calculations for various device thicknesses.**

Energy–phase dispersion of subgap states $E(\phi)$ for a weak link device with four different thicknesses: (**a**) 120 nm, (**b**) 180 nm, (**c**) 220 nm, and (**d**) 320 nm. Each thickness shows multiple Andreev bound-state branches originating from the involved transport channels. All spectra are calculated self-consistently within the Wannier-based BdG framework.

We now relate these ABS spectral features to the supercurrent response by examining the current–phase relation (CPR). We calculate the equilibrium supercurrent—Josephson current, from the phase derivative of the BdG free energy (**eq. 19**). Because the supercurrent is directly proportional to the derivative of the ABS energies with respect to phase, changes in thickness affect the spectral slopes $dE_n/d\phi$. We observe significant variations in CPR amplitude and shape with thickness (**Fig. 7**). Even though the critical supercurrent ($I_c = \max_{\phi} | I(\phi) |$

) remains in the nA range across all studied thicknesses, the CPR significantly deviates from a simple sinusoidal form. We can quantify this deviation using a harmonic expansion given by $I(\phi) = a_1 \sin\,\phi + a_2 \sin\,(2\phi) + \cdots$. Here, the ratio $a_2/a_1$ measures the skewness of the CPR, where a nonzero second harmonic shifts the phase ($\phi_{\max}$) away from $\pi/2$, at which the supercurrent is maximum ($I_c = \max_{\phi} \mid I(\phi) \mid$) [58-60].

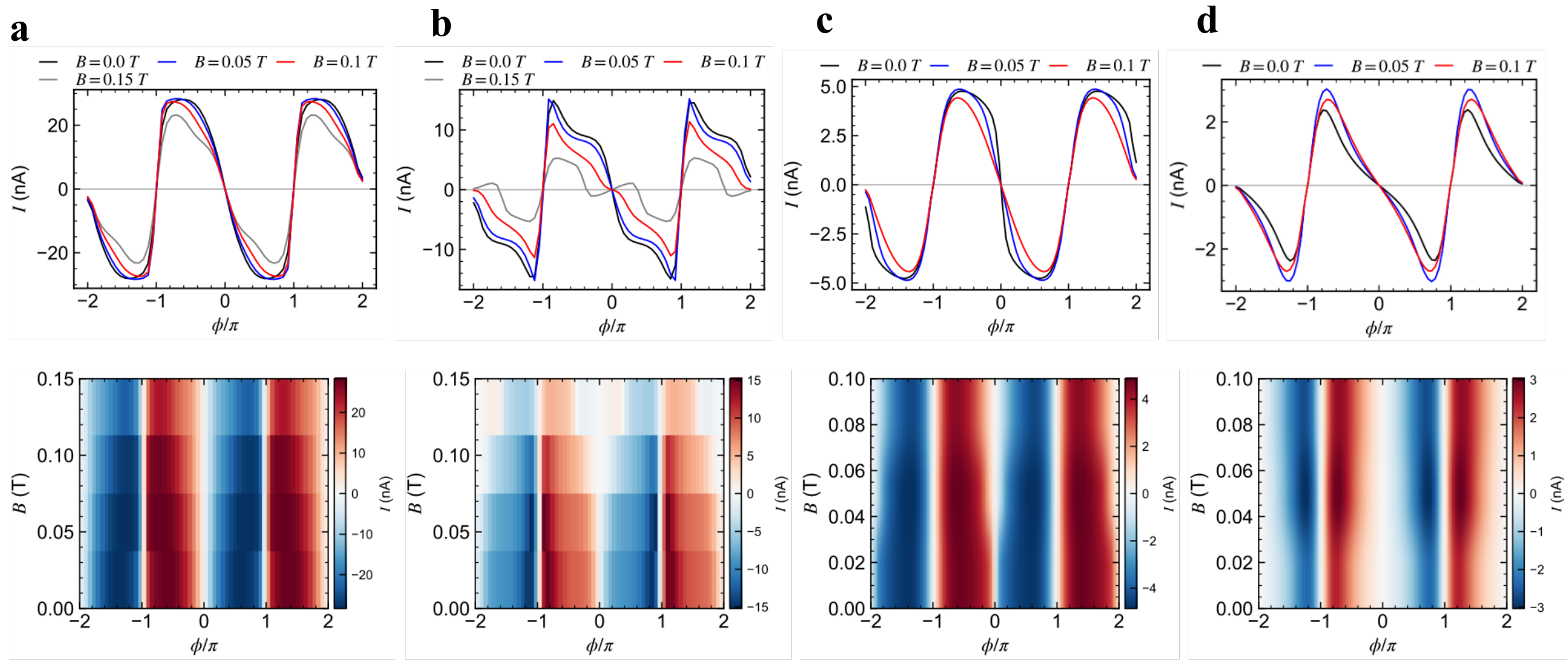

**Figure 7. Calculated current–phase relation (CPR) of the Nb—1D $Bi_4Br_4$—Nb weak link.**

Josephson supercurrent $I(\phi)$ as a function of the phase difference $\phi$ of the Nb—1D $Bi_4Br_4$—Nb weak link with different link sizes: (**a**) 120 nm, (**b**) 180 nm, (**c**) 220 nm, and (**d**) 320 nm, under various magnetic fields. The bottom panel shows heat maps of the supercurrent versus phase at different magnetic fields. CPR was calculated from the full Bogoliubov–de Gennes spectrum, including the spin-orbit coupling (SOC) effect.

The calculated skewness value of +1.74 indicates that the CPR of the Nb—1D $Bi_4Br_4$—Nb device exhibits a forward skewing feature, which physically suggests a transition from a standard sinusoidal (i.e., junction with low transmission coefficient) junction to a nearly ballistic (i.e., junction with large transmission coefficient) junction. Such skewing will significantly enhance the junction's sensitivity to magnetic fields, producing sharper, better-defined signals. Consequently, this weak link device is expected to perform well in detecting small magnetic field variations, such as in SQUID applications. Importantly, this CPR can exhibit such a second-harmonic shift even while remaining odd and preserving particle–hole symmetry. However, this skewness does not indicate a true anomalous junction, as it would require a finite cosine component and broken inversion symmetry. Instead, it reflects the nontrivial phase dispersion relation of multiple ABS in the intermediate-to-long-junction regime.

Intriguingly, the external magnetic field significantly tunes the CPR with only a minor change in its overall amplitude (top panel of **Fig. 7**). Since the normal region in the Nb—1D $Bi_4Br_4$—

Nb device is periodic only along the transport direction ($x$), with vacuum in the other transverse directions, the orbital flux through the weak link should be negligible. Therefore, the primary microscopic interaction is Zeeman splitting alone, which is expressed by $H_x = (g\mu_B/2)\,\mathbf{B}\cdot\boldsymbol{\sigma}$. This magnetic field distorts the ABS spectrum by shifting avoided crossings and lifting spin degeneracies. As a result, the magnetic field mainly affects the shape of the CPR rather than the magnitude of the supercurrent ($I_c$). Consequently, the critical supercurrent shows only a slight dependence on the magnetic field, while the phase at which the current peaks remains largely unaffected. Meanwhile, the local slope $dI/d\phi$ varies significantly with external magnetic field B (e.g., from approximately 4 to 7 nA/rad). As clearly visible in the top panel of **Fig. 7**, increasing the magnetic field causes noticeable phase shifts and changes in skewness for each thickness, even though the peak supercurrent changes only marginally.

The phase sensitivity of the studied device to the magnetic field is further demonstrated by the heat maps at the bottom of **Fig. 7**. Although the response to the magnetic field is not uniform, it varies significantly with phase. We observe the highest value of $|\,\partial I/\partial B\,|$ at specific phases. At these points, the ABS spectra are most sensitive to $\phi$ and the Zeeman field splitting. For the weak link device examined here, the optimal operating point provides a maximum $|\,\partial I/\partial B\,|$ of around 6 nA/T, where

$$\frac{\partial I}{\partial B} = \frac{2e}{\hbar}\sum_{n>0}\left[\frac{\partial E_n}{\partial \phi}\,\frac{\partial}{\partial B}\tanh\left(\frac{E_n}{2k_BT}\right) + \frac{\partial^2 E_n}{\partial B\,\partial \phi}\tanh\left(\frac{E_n}{2k_BT}\right)\right],$$

Although the change in supercurrent value with different magnetic fields is minimal, it underscores an important design principle for superconducting sensors. In these sensors, magnetic sensitivity relies less on the supercurrent's amplitude and more on the local phase sensitivity of CPR to the magnetic field. In practical devices, this phase sensitivity can be measured by exploiting the Josephson inductance $L_J(\phi, B) = \frac{\Phi_0}{2\pi}\left(\frac{dI}{d\phi}\right)^{-1}$ [61] [62]. This physical property can translate phase sensitivity into a measurable signal. Overall, this systematic study—combining thickness tunability, CPR skewness, and phase-sensitive magnetic sensitivity—establishes Nb—1D $Bi_4Br_4$—Nb as a versatile platform for phase-coherent device engineering and for compact, low-current magnetic sensing applications.

## Conclusion

We computationally design an Nb—1D $Bi_4Br_4$—Nb weak link device and investigate its transparency and superconducting transport properties using first-principles calculations, a tight-binding Hamiltonian, and BdG equations. The electronic structure of bulk Nb confirms its metallic nature, while phonon calculations show a strong electron–phonon coupling strength (1.19), leading to a superconducting transition temperature of about 9 K, which matches well with known experimental data for Nb. In contrast, the topological $Bi_4Br_4$ nanoribbon exhibits metallic edge states with topological protection but does not display any intrinsic superconductivity. As a result, the Nb—1D $Bi_4Br_4$—Nb weak link acts as an S–N (with topological protection)–S junction, where the proximity effect of Nb induces superconductivity only.

We identify a limited number of transport channels at this weak link from normal-state transport calculations. Specifically, two counterpropagating channels have nearly 0.86 transmission probabilities, while two other channels show moderate transmission. However, other higher modes are effectively closed, i.e., not transmitting. This transparency characteristic places the Nb—1D $Bi_4Br_4$—Nb junction in a high-transparency, few-mode regime. We also demonstrate Andreev process–dominated superconducting transport within the subgap region from differential conductance calculations. Additionally, we reveal multichannel Andreev bound states (ABS) with strong phase dispersion via full Bogoliubov–de Gennes (BdG) calculations, owing to nearly ballistic transparency. This strong phase dispersion of the ABS results in a non-sinusoidal current–phase relation (CPR) with significant forward skewness (+1.74) and high phase sensitivity to magnetic fields. Overall, these findings show that the Nb—1D $Bi_4Br_4$—Nb device hosts robust, high-transmission channels and exhibits strong proximity-effect–induced superconductivity, making this platform promising for compact superconducting Josephson devices and magnetic-field sensing applications.

**Acknowledgement**

This work was financially supported with a grant CNS2023-144994 funded by MICIU/AEI/10.13039/501100011033 and by "ERDF/EU."

**Conflict of Interest**

There is no conflict of interest to declare.

**Data Availability**

All data are included in the manuscript, and the raw data are available upon request from the corresponding author.